# Earthquake Occurrences in the Pacific Ring of Fire Exhibit a Collective Stochastic Memory for Magnitudes, Depths, and Relative Distances of Events

**Pamela Jessica C. Roque**[1*], **Renante R. Violanda**[1], **Christopher C. Bernido**[1,2], **Janneli Lea A. Soria**[3]

[1]*Department of Physics, University of San Carlos, Talamban, Cebu City 6000, Philippines*

[2]*Research Center for Theoretical Physics, Central Visayan Institute Foundation, Jagna Bohol 6308, Philippines*

[3]*JAZC Marine Sciences Laboratory, Central Visayan Institute Foundation, Jagna Bohol 6308, Philippines*

* Correspondence:
Pamela Jessica C. Roque
644014@usc.edu.ph

**Keywords:** Earthquakes, Pacific Ring of Fire, Stochastic process with memory, Mean square displacement, Earthquake catalog**.**

**Abstract**

Around 90% of the world's earthquakes occur at the circum-Pacific belt referred to as the Pacific Ring of Fire exposing the countries in this region to high risk of earthquake hazards. We model fluctuations of the different seismic magnitudes, interevent distances, and seismic depths as a function of earthquake occurrence from the earthquake catalogs of Chile, Mexico, Japan, New Zealand, Philippines, and Southern California as a stochastic process with long-term memory. We show that the fluctuations of the three seismic quantities mentioned for all regions studied in this paper are governed by a single memory function that is described by a memory parameter $\mu$ and a decay parameter $\beta$. The values of $\mu$ exhibit an underlying characteristic memory behavior of seismic activities common to all the countries considered, while the values of $\beta$ suggest a regional dependence which could be a manifestation of different seismic dynamics in various regions. This new perspective may provide a more versatile approach in studying the independent datasets that may be extracted from various earthquake catalogs.

## 1  Introduction

The spatiotemporal complexity of seismicity and its stochastic nature are the main reasons why studying earthquakes continue to be a great scientific challenge. The apparent randomness of earthquake occurrences has made it difficult, if not impossible, to produce accurate hazard assessments or earthquake forecasts (King et al. 1994; Felzer et al. 2002; Felzer et al. 2003; Kagan and Jackson 2014; De Arcangelis et al. 2016; Fan et al. 2019; Hardebeck 2020). Advancement in instrumentation however paved the way for the development of comprehensive earthquake catalogs containing the location, magnitude, occurrence time, and depth of seismic events. The availability of earthquake catalogs has allowed analysis of earthquakes and earthquake sequences, particularly their emergent statistical nature.

An alternative approach to studying individual earthquake events and sequences is through scaling relationships and statistical correlations based on interevent spatiotemporal distributions of earthquakes (Corral 2003; Corral 2004; Fan et al. 2019; Zhang et al. 2020; Zhang et al. 2021a). Perhaps the most popular distribution characterizing earthquakes is the power-law decrease of earthquakes in a specific region with energies greater than a specific threshold. This is the well-known Gutenberg-Richter (G-R) law of earthquake magnitudes, which is a hallmark of earthquakes' scale-free behavior (Gutenberg and Richter, 1944). G-R law states that earthquake magnitude distribution follows a power law trend past a magnitude threshold $M_c$. For earthquake magnitudes less than $M_c$, the catalog is considered incomplete, i.e., the catalog is complete for magnitudes $\geq M_c$ (Batac and Kantz 2014; Zhang et al. 2021a,b). Furthermore, the G-R distribution is preserved at any given region (Gutenberg and Richter, 1944), i.e., G-R law is generalizable in space. It was also found that the spatial distribution of epicenters is fractal over the surface of the Earth (Turcotte, 1997). The Omori Law, which accounts for the number of events (called aftershocks) which follow a large earthquake after some time likewise decays as a power-law. As the name suggests, scale-free distributions imply the lack of characteristic scales which define the seismic process. Nevertheless, recent studies on inter-event statistics, (Batac and Kantz, 2014; Batac 2016; De Arcangelis et al. 2016; Wang et al. 2017) and conditional probabilities (Touati et al. 2009; Zhang et al. 2020) reveal vital correlations between successive events which pave the way for more functional means of analyzing seismicity (Baiesi and Paczuski, 2004). Specifically, studies have shown that earthquakes and earthquake clusters are processes with memory (Fan et al. 2019; Zhang et al. 2020; Gkarlaouni et al. 2017; Livina et al. 2005). In areas where there is interaction between tectonic faults, memory effect also appears to be strong (Gkarlaouni et al. 2017).

Systems with memory have been modeled using the Hida white noise functional integral (Hida et al. 1993; Bernido & Carpio-Bernido 2012, 2015) approach where the evolution of a fluctuating observable is mathematically parametrized to uncover the inherent non-Markovian stochastic process. This analytical tool has been successful in analyzing various complex systems such as the degradation of the Great Barrier Reef, sea surface temperatures, atmospheric carbon dioxide levels (Elnar et al. 2021), fibrin ageing (Aure et al. 2019), nucleotide distribution in genomes (Violanda et al. 2019), diffusion coefficient in biomolecular transport (Barredo et al. 2018), and tropical cyclone track (Bernido et al. 2014), among others. In this work, three fluctuating observables contained in an earthquake catalog: earthquake depths, magnitudes, and interevent distances are modeled using the Hida stochastic functional integral technique. Particularly, we extract the fluctuations from the local catalogs of six different regions in the Pacific Ring of Fire where seismic activities are highest (Figure 1). The mean square displacements (MSD) from the empirical fluctuations are matched with a theoretical MSD which allows the determination of the parameter values of the stochastic model. The exact values of the parameters are used to generate the probability density functions (PDF) for each dataset. We show for the first time that earthquakes in six countries around the Pacific Ring of Fire exhibit a single collective memory, which conforms with the white noise functional integral model. This study demonstrates that the stochastic framework can be used to extract patterns in seemingly random data series such as earthquakes.

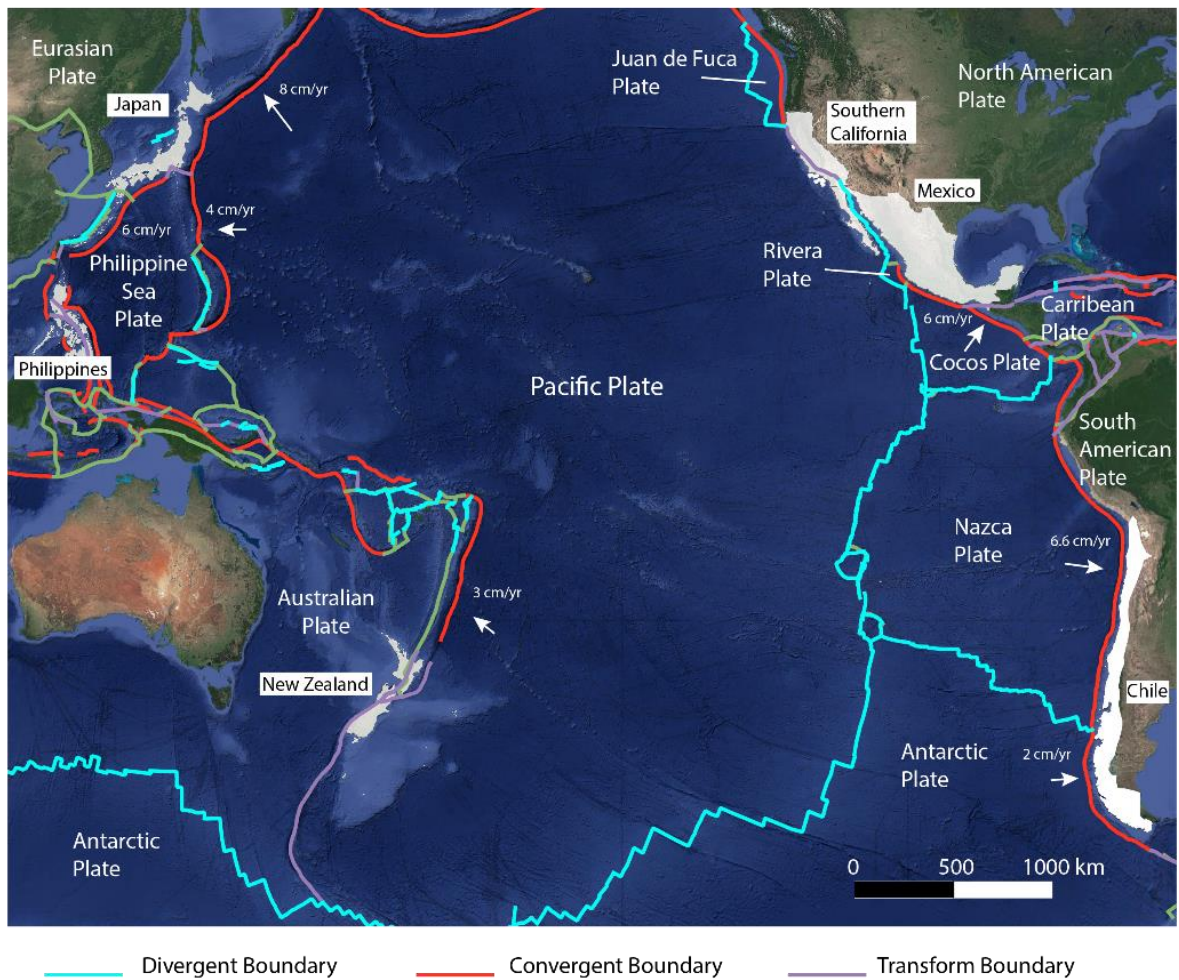

**Figure 1. Regions of study, highlighted in white, along the Pacific Ring of Fire.**

## 2 Materials and methods

### 2.1 Local earthquake catalogs

Earthquake data from six regions across the Pacific Ring of Fire, mostly covering the time periods from 1990s to 2020s were requested or accessed through the respective online repositories (Table 1). Using the earthquake datasets corresponding to each region, the distances of epicenters between consecutive earthquakes are calculated. The other datasets modeled in this paper are the hypocenter depths and magnitudes of successive earthquake events.

**Table 1. Information on the regional earthquake catalogs used in this study.**

| Region | Earthquake catalog source | Years of coverage | Events count |
|---|---|---|---|
| Chile | Centro Sismológico Nacional (CSN) | 2000 to 2021 | ~117,000 |
| Japan | Japan Meteorological Agency (JMA) | 1999 to 2019 | ~2,568,000 |
| Mexico | Servicio Sismológico Nacional (SSN) | 1994 to 2020 | ~186,000 |
| New Zealand | GeoNet | 1994 to 2020 | ~525,000 |
| Philippines | Philippine Institute of Volcanology and Seismology Seismological Observation and Earthquake Prediction Division (PHIVOLCS) | 1994 to 2020 | ~88,000 |
| Southern California | Southern California Earthquake Data Center (SCEDC) | 1994 to 2020 | ~515,000 |

## 2.2 Distances between consecutive earthquake epicenters

The distance between consecutive earthquakes, denoted by $\Delta r_{ij}$, with epicenters at $r_i$ and $r_j$ can be calculated as an arc length on the surface of the Earth, known as the Haversine Law (Batac and Kantz, 2014). The distance $\Delta r_{ij}$ is given by,

$$\Delta r_{ij} = E\cos^{-1}[\sin\theta_i \sin\theta_j + \cos\theta_i \cos\theta_j \cos(\varphi_i - \varphi_j)] \qquad (1)$$

where distance $E$ = 6371 km is the radius of the Earth. In our treatment of the earthquake data, the first event in the data series corresponds to an occurrence number $\tau = 0$. The indices $i$ and $j$ represent consecutive earthquake events where event $i$ is defined as an event with occurrence $(t + \tau)$ while the occurrence number for event $j$ is $(t)$. The sequence of earthquake events will then be numbered consecutively by an occurrence number. Figure 2 shows the series of distance measurements, ranging from a few meters to thousands of kilometers apart, of consecutive earthquakes for the different data sets used in this study.

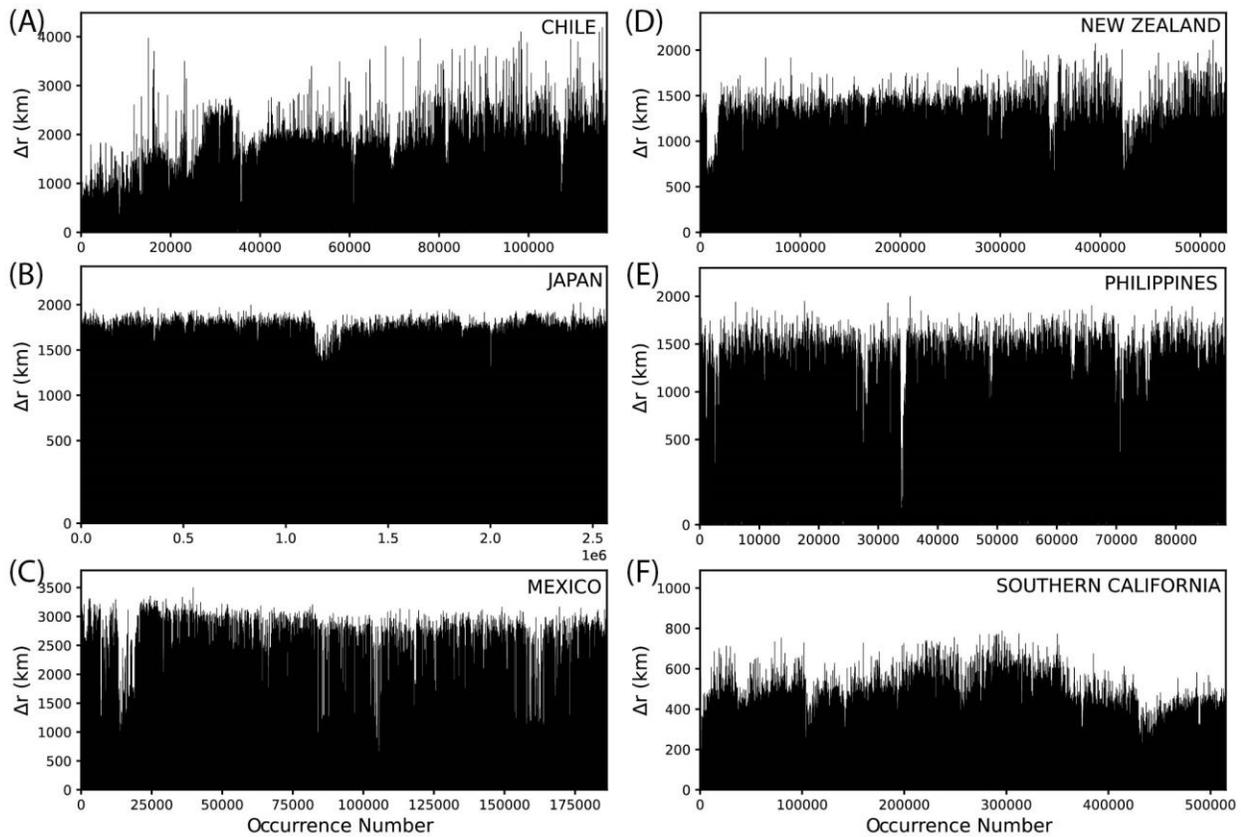

**Figure 2.** The series of distances between consecutive earthquake epicenters for all events in each region.

## 2.3 Hypocenter depths

Hypocenter depth distributions reveal variations in rheology of the Earth's crust which govern the temporal changes in the Earth's continents and the current tectonic processes (Sloan et al. 2011). In this study, another dataset that will be shown to manifest a stochastic process with memory is the series of consecutive earthquake depths. Each earthquake is assigned to an occurrence number which follows the chronological sequence of events as shown in Figure 3. Similar to earthquake epicenter distances, the dataset on earthquake depths is used to validate the stochastic model.

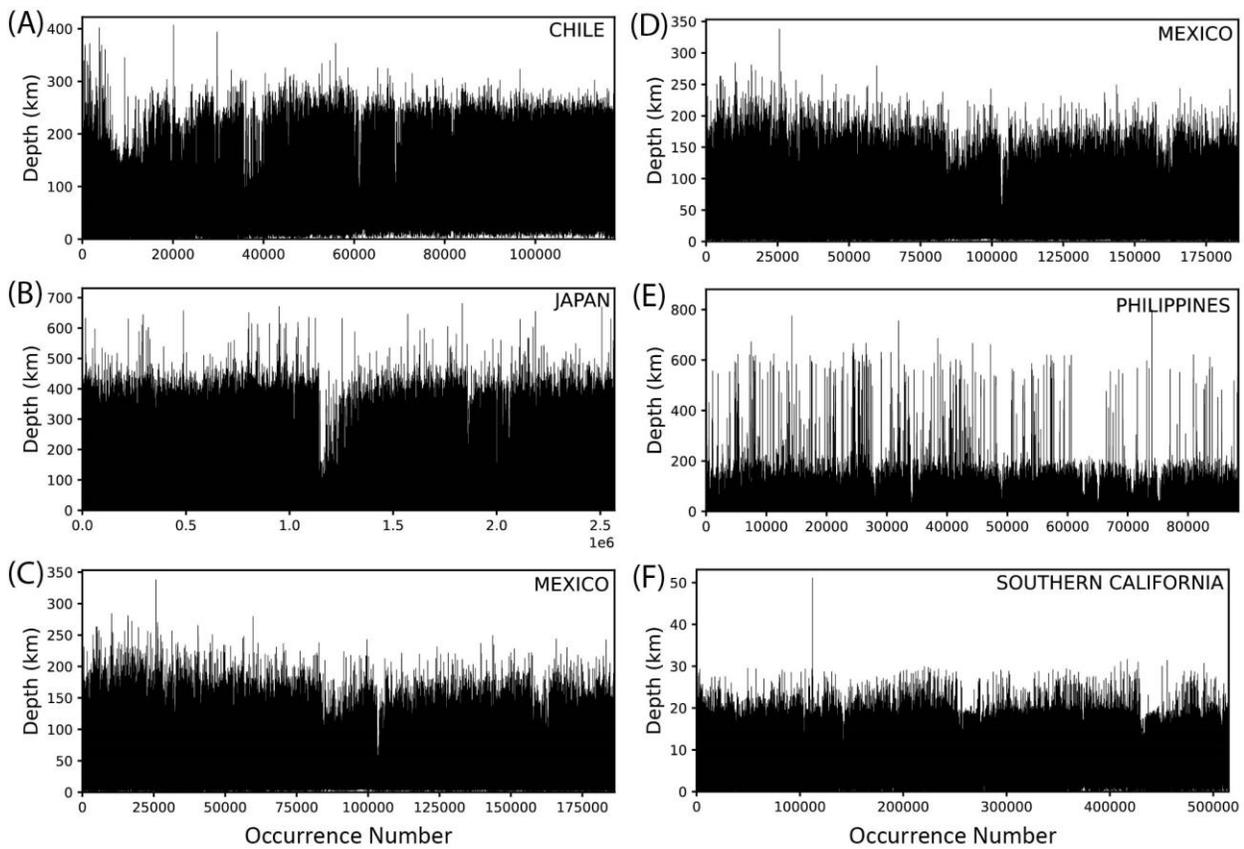

**Figure 3.** Hypocenter depth series for all events in each region.

## 2.4 Earthquake magnitudes

To further characterize seismicity, the fluctuating earthquake magnitudes are also taken into account. Similar to the first two datasets, each earthquake is labelled by an occurrence number that preserves the actual sequence of events as shown in Figure 4. The empirical MSD is then used to test the stochastic model representing fluctuations in systems with memory.

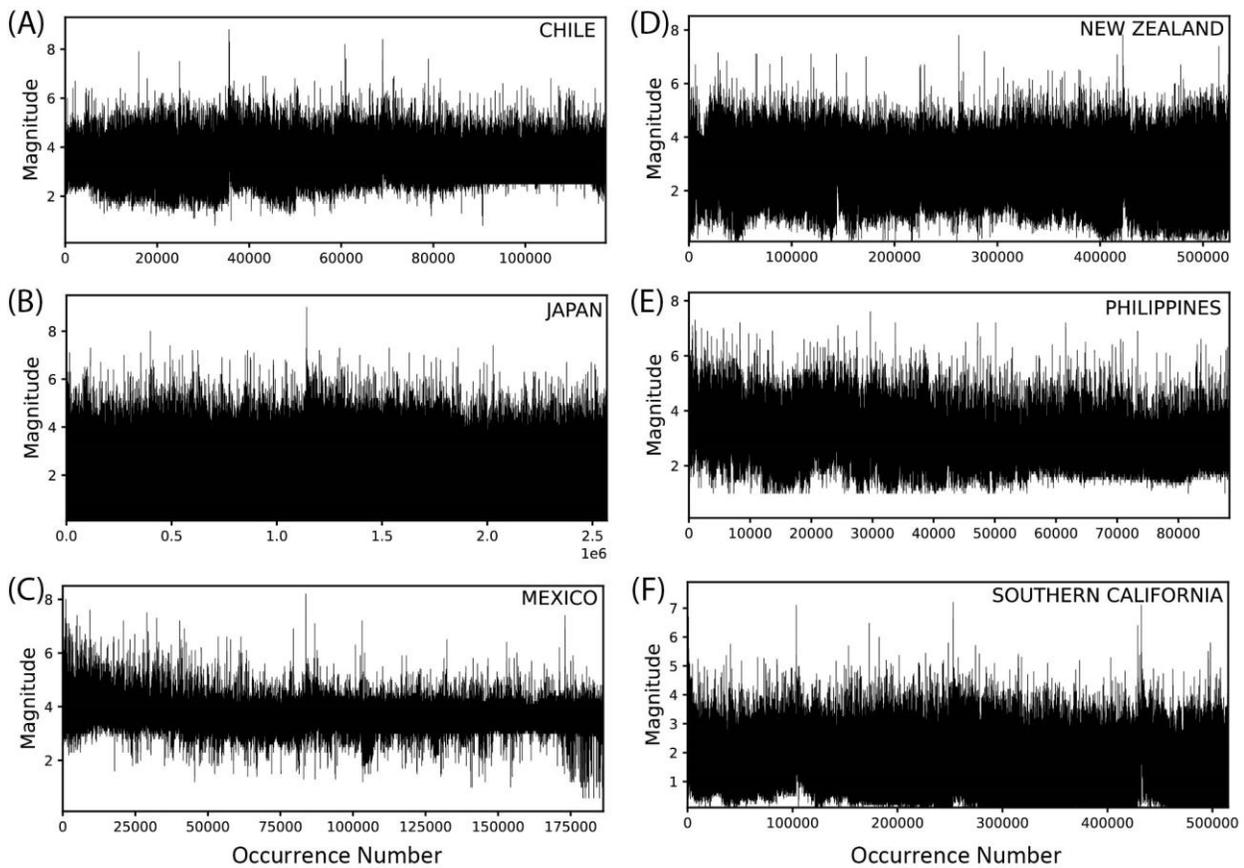

**Figure 4. Series of earthquake magnitudes for all events in all regions.**

To test the generalizability of the theoretical MSD function in event or magnitude completeness, it is applied to partitioned seismic catalogs. The subsets of an earthquake catalog are obtained by setting various cutoff magnitudes, which verifies the contribution of completeness magnitude to the robustness of the derived functions. Figure 5 shows the series of epicenter distance measurements, depths, and magnitudes of consecutive earthquakes with cutoff magnitudes of 2.0, 3.0, 4.0, and 6.0. It can be seen that at a larger cut-off frequency, the data becomes sparse. Despite this, there is only little variation in the range of distance values: consecutive epicenter distances can be a few meters to thousands of kilometers apart.

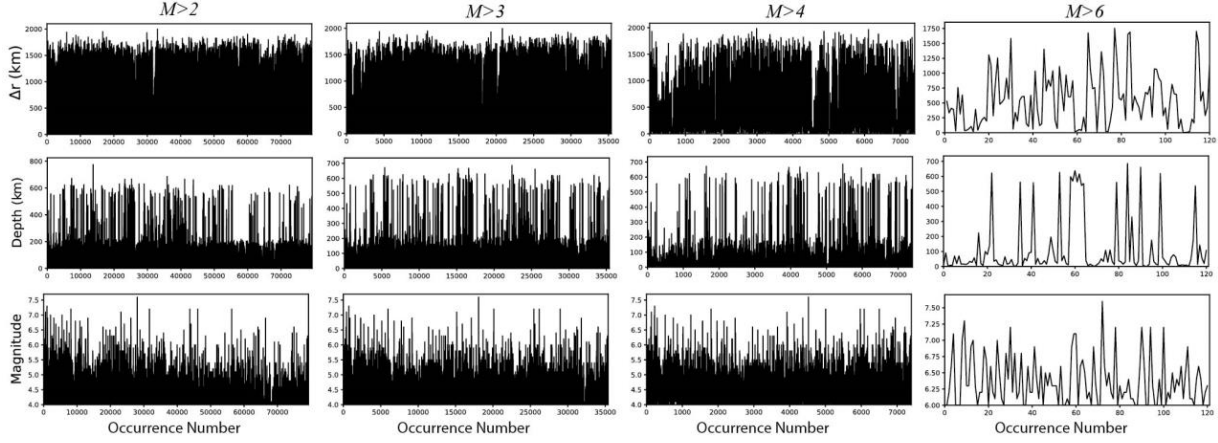

**Figure 5.** The series of Philippine earthquake interevent distances, depths, and magnitudes for events with magnitudes 2.0, 3.0, 4.0, and 6.0.

## 2.5 Stochastic framework with memory

For earthquake events, we write the fluctuating observable as $x(t)$, i.e., $x(t) = x_0 + fluctuations$, where $t$ is time and, $x(0) = x_0$, is the initial value. In particular, the stochastic fluctuations are parametrized in terms of the Brownian motion $B(\tau)$ as,

$$x(t) = x_0 + \int_0^t (t-\tau)^{\frac{(\mu-1)}{2}} e^{\frac{-\beta}{2\tau}} \tau^{\frac{-(\mu+1)}{2}} dB(\tau) \qquad (2)$$

where $0 \leq \tau \leq t$. In Equation (2), $x(t)$ models the fluctuating earthquake interevent distances, magnitudes, or depths from the designated initial occurrence $\tau = 0$ to occurrence at, $\tau = t$. Note that the factor, $(t-\tau)^{\frac{(\mu-1)}{2}}$ acts as a memory kernel describing the physical system and $e^{\frac{-\beta}{2\tau}} \tau^{\frac{-(\mu+1)}{2}}$ modulates the Brownian fluctuation $B(\tau)$. The $\beta$ and $\mu$ are parameters to be determined from the dataset.

The probability density function (PDF) corresponding to Equation (2) is given by (see Supplemental Material, Note S2),

$$P(x_1, t; x_0, 0) = \frac{1}{\sqrt{2\pi\Gamma(\mu)\beta^{-\mu}t^{\mu-1}e^{-\beta/t}}} exp\left(\frac{-(x_1-x_0)^2}{2\Gamma(\mu)\beta^{-\mu}t^{\mu-1}e^{-\beta/t}}\right) \qquad (3)$$

where $x_1$, is the value of the fluctuating variable at final time $t$, i.e., $x(t) = x_1$.

With Equation (3), one could also obtain the mean square displacement (MSD) given by, $\text{MSD} = \langle x^2 \rangle - \langle x \rangle^2$, where $\langle x^2 \rangle = \int_{-\infty}^{+\infty} x^2\, P(x,t;x_0,0)\, dx$, with $P(x,t;x_0,0)$ given by Equation (3) (Bernido and Carpio-Bernido, 2015). This yields,

$$MSD = \Gamma(\mu)\beta^{-\mu}t^{\mu-1}e^{-\beta/t} \qquad (4)$$

To determine the $\beta$ and $\mu$ parameters, we match the theoretical MSD, Equation (4) with the empirical MSD. Designating $x$ as the value of the observable in a dataset at a particular occurrence number $i$, the empirical MSD is calculated using,

$$MSD(\Delta) = \frac{1}{N-\Delta}\sum_{i=1}^{N-\Delta}[x(i+\Delta) - x(i)]^2 . \quad (5)$$

Here, $N$ is the total number of data points with, $\Delta < N$, as the occurrence interval. The occurrence interval $\Delta$ is the interval or separation between two occurrence numbers. An increasing $\Delta$ generates the plot for the empirical MSD.

## 3  Results and discussion

### 3.1  Empirical and theoretical MSD

We first consider the fluctuating values for the series of distances between consecutive earthquake epicenters shown in Figure 2. The MSD of these fluctuations is plotted as a function of the occurrence interval $\tau$ which is used to validate the analytical model described by Equation (4).

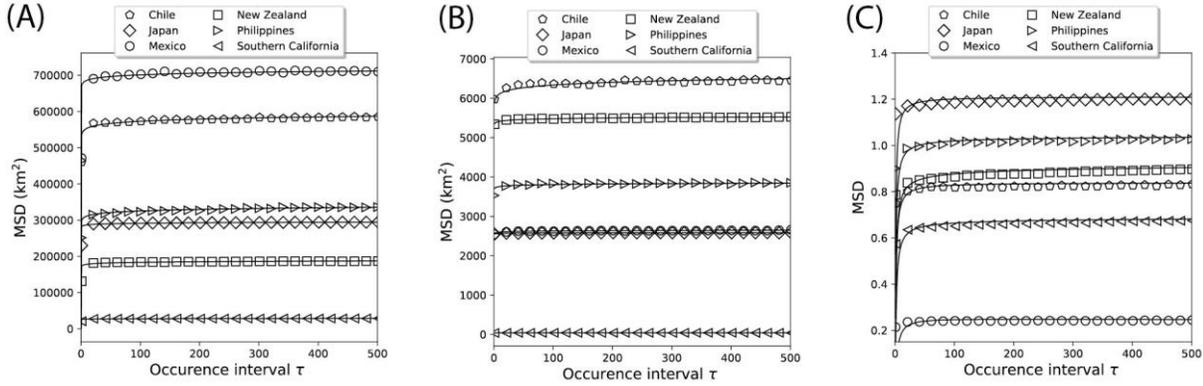

**Figure 6. MSD's of earthquake (A) epicenter distances, (B) depths, and (C) magnitudes for the entire data set in each catalog. Symbols are from the empirical data and solid lines are theoretical MSD's from Equation (4).**

Figure 6 shows the plot for the MSD of different regions based on the dataset shown on Figures 2-4. The black solid line in Fig. 6 is the theoretical MSD, Eq. (4). For the different regions, the theoretical MSD in Fig. 6 generally captures the shape of the empirical MSD. In Figure 6A, the variation of the values of the parameters $\beta$ and $\mu$ across regions ranges as follows: $\beta$ changes from $1.68 \times 10^{-6} - 4.48 \times 10^{-5}$; and $\mu$ changes from 1.0068 to 1.0191 (see, e.g., Supplemental Material, Note S1).

Figure 6B displays the respective MSD of the consecutive hypocenter depths for various regions shown in Figure 3. Similar to the interevent distances, Eq. (4) also provides a good match for the empirical data of the consecutive hypocenter depths. The values of the parameters $\beta$ and $\mu$ vary between the ranges $1.92 \times 10^{-4} - 2.82 \times 10^{-2}$, and $1.002 - 1.015$, respectively.

To further unravel the complexity of earthquakes, the magnitudes of consecutive earthquakes are also analyzed. Based on the fluctuating values in Figure 4, we plot the MSD versus occurrence interval $\tau$ in Figure 6C for the series of earthquake magnitudes for all events recorded within the time span considered for each region. The shape of the empirical MSD's is captured by the same analytical MSD, Equation (4). For Equation (4) to match the empirical MSDs in Figure 6C, the parameter values for $\beta$ and $\mu$ are observed to change within the following boundaries: $\beta$ varies from 0.83 to 3.74; and $\mu$ varies from 0.9835 to 1.0191.

The MSD's shown in Figures 6 are characterized by a short initial period in which the values rapidly change followed by a mild increase at the end. The exact values of $\beta$ and $\mu$ for each dataset and region are listed in Table 2.

**Table 2.** Parameter values for Figures 6-8.

|  | $\beta$ | | | $\mu$ | | |
| --- | --- | --- | --- | --- | --- | --- |
|  | Epicenter | Depth | Magnitude | Epicenter | Depth | Magnitude |
| Chile | $(2.250 \pm 0.05) \times 10^{-6}$ | $(1.916 \pm 0.003) \times 10^{-4}$ | $1.195 \pm 0.003$ | $1.0149 \pm 0.0001$ | $1.0154 \pm 0.0001$ | $1.0004 \pm 0.0005$ |
| Japan | $(3.839 \pm 0.007) \times 10^{-6}$ | $(3.9728 \pm 0.0009) \times 10^{-4}$ | $0.825 \pm 0.002$ | $1.00676 \pm 0.00009$ | $1.00193 \pm 0.00002$ | $0.9998 \pm 0.0003$ |
| Mexico | $(1.679 \pm 0.005) \times 10^{-6}$ | $(4.014 \pm 0.003) \times 10^{-4}$ | $3.74 \pm 0.02$ | $1.0094 \pm 0.0002$ | $1.00444 \pm 0.00005$ | $0.9835 \pm 0.0009$ |
| New Zealand | $(6.69 \pm 0.02) \times 10^{-6}$ | $(1.947 \pm 0.001) \times 10^{-4}$ | $1.187 \pm 0.003$ | $1.0129 \pm 0.0001$ | $1.00516 \pm 0.00004$ | $1.0129 \pm 0.0005$ |
| Philippines | $(4.21 \pm 0.01) \times 10^{-6}$ | $(2.813 \pm 0.004) \times 10^{-4}$ | $0.990 \pm 0.002$ | $1.0191 \pm 0.0001$ | $1.00563 \pm 0.00009$ | $1.0191 \pm 0.0004$ |
| Southern California | $(4.482 \pm 0.008) \times 10^{-5}$ | $(2.822 \pm 0.001) \times 10^{-2}$ | $1.529 \pm 0.005$ | $1.0157 \pm 0.0001$ | $1.01202 \pm 0.00004$ | $1.0157 \pm 0.0005$ |

## 3.2 Robustness of MSD function

For different cutoff magnitudes, the comparison of the theoretical and empirical MSD's can also be done. Magnitude cutoffs can be related to earthquake magnitude catalog completeness, as described in the Gutenberg-Richter law. Based on Figure 5 (different cutoff magnitudes from 1994-2020, Philippine earthquake catalog), we show the theoretical and empirical MSD comparison for interevent distances, earthquake depths, and magnitudes in Figures 7A to 7C, respectively. The solid line in Figure 7 is the theoretical MSD, Equation (4). As the cut-off magnitude is increased, the values of the MSD increase for the epicenter distances as shown in Figure 7A. This signifies that, on average, the distance among stronger earthquakes is larger, which may be attributed to the removal of aftershock/foreshock sequences, since these events cluster in space as shown by Batac and Kantz (2014) and Zhang et al (2020). Figure 7B shows that the MSD of the hypocenter depths are also observed to increase as the lower magnitude seismic events are removed. This indicates that sequences of higher magnitude earthquakes do not cluster in depth and are likely to be independent earthquake events. For different earthquake magnitude cutoffs, the theoretical MSD in Figure 7C still captures the shape of the empirical MSD, regardless of the set magnitude cutoff. Setting higher cutoffs constrain the magnitude values which results in a lower MSD. These results show that Equation (4) describing the three independent datasets is insensitive to magnitude completeness in seismic catalogs. Similar patterns are observed from the earthquake datasets of the other regions. The exact values of $\beta$ and $\mu$ are summarized in the Supplemental Material.

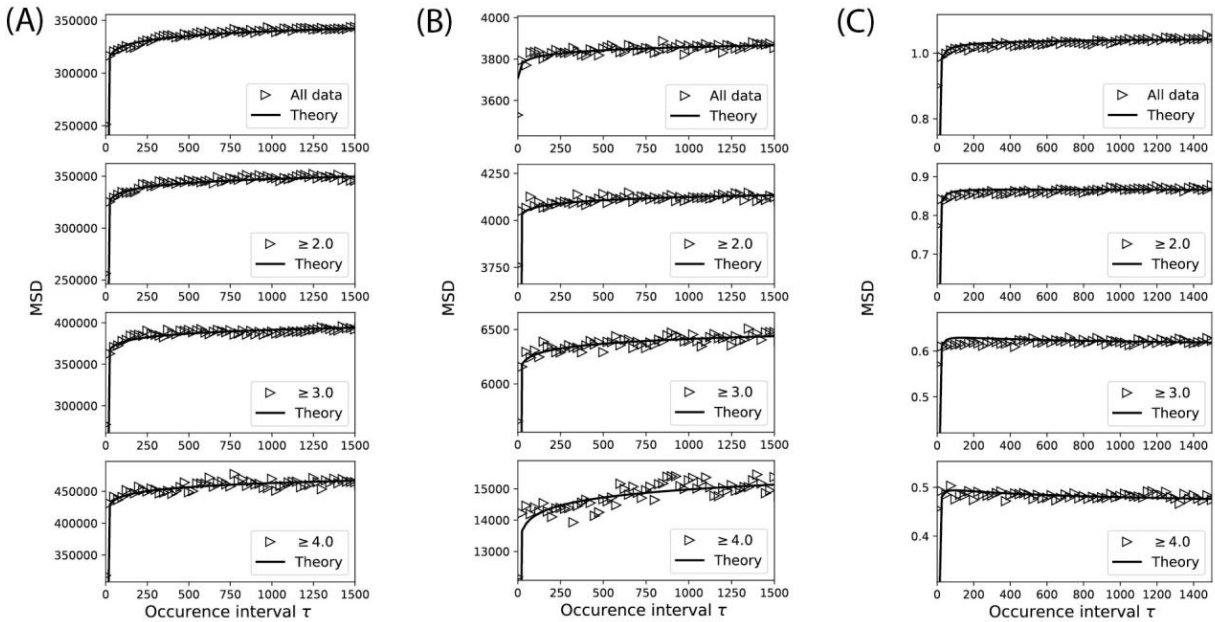

**Figure 7. MSD of earthquakes for (A) interevent distances, (B) depths, and (C) magnitudes, for various cutoff magnitudes from PHIVOLCS earthquake catalog. Black solid lines are the theoretical MSD using Equation (4).**

### 3.3 Probability density function

The theoretical probability density function (PDF), Equation (3), can be tested against the empirical dataset. Using the $\beta$ and $\mu$ values obtained from matching the empirical and theoretical MSD's, the probability distributions for different magnitude cutoffs as a function of displacement values for a fixed occurrence interval can also be generated.

### 3.3.1 Displacement distribution

We present, as an example, a detailed analysis using the PHIVOLCS earthquake catalog. Similar results are obtained for the datasets of the other regions. The theoretical PDF for a given seismic event interval $\tau$ is compared with the dataset for displacement distribution, $\Delta x = (x(i + \Delta) - x(i))$, where $x$ can be the earthquake epicenter distance, depth, or magnitude. The results for the PDF are shown for interevent distances (Figure 8A), depths (Figure 8B), and magnitudes (Figure 8C), respectively. All figures show a peak near $\Delta x = 0$.

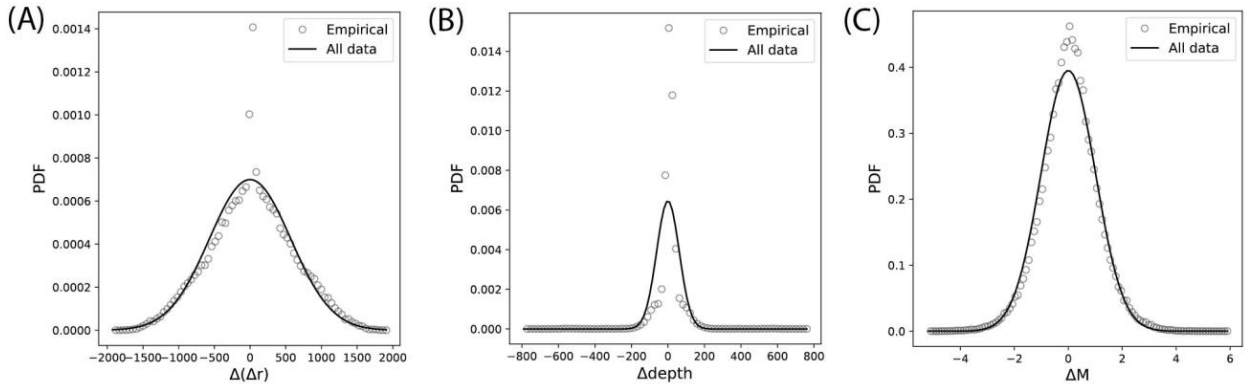

**Figure 8. PDF's of earthquake (A) interevent distances, (B) depths, and (C) magnitudes in the Philippines for all data in the catalog.**

### 3.3.2 Earthquakes with magnitude ≥ 6.0

Earthquakes with magnitudes of 6 or greater can cause significant damage and are of special interest in the context of disaster preparedness. With much fewer data for magnitudes $\geq 6.0$, we show in Figure 9, the theoretical PDF for $\tau = 3$, compared with that from the dataset of different regions.

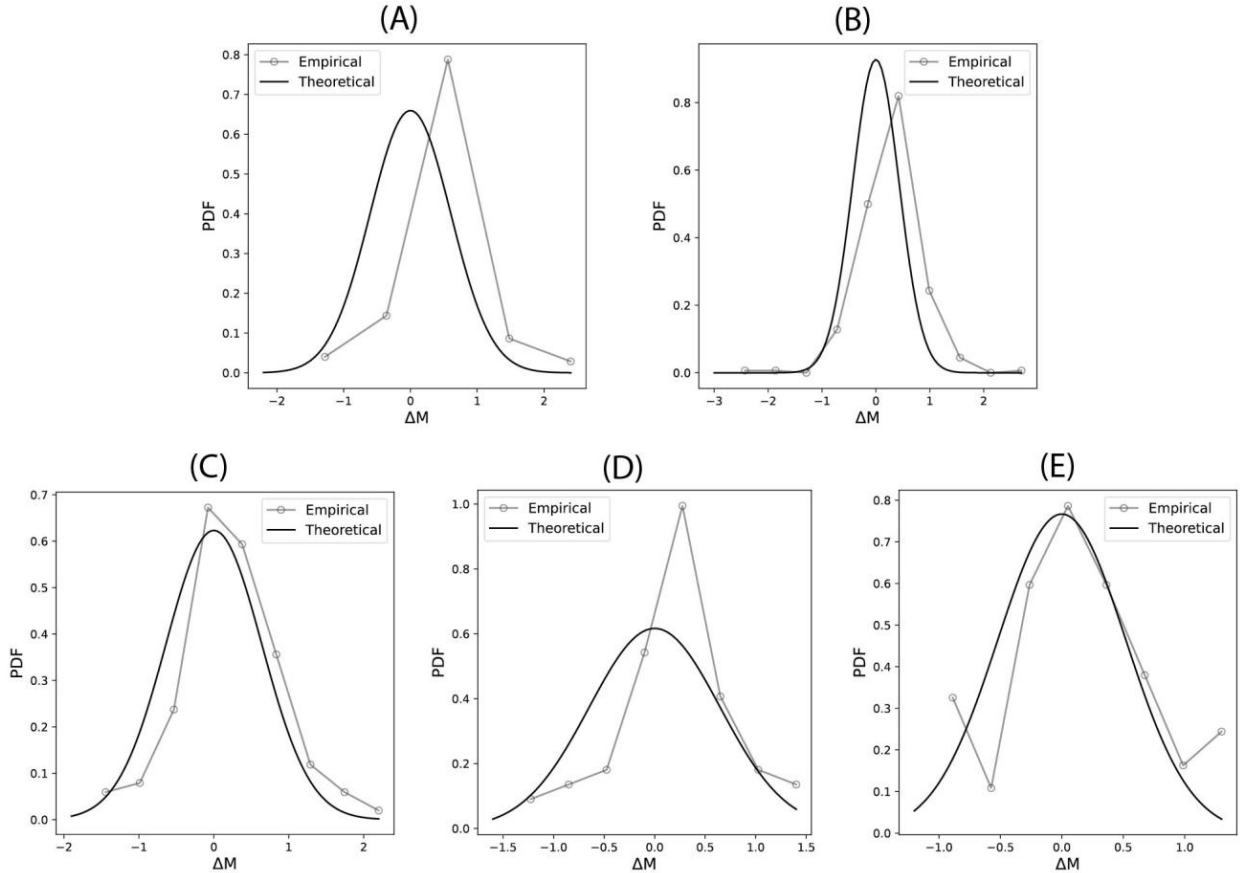

**Figure 9.** Earthquake of magnitudes ≥6.0. Theoretical PDF (τ=3) matched with the dataset from different countries: (A) Chile; (B) Japan; (C) Mexico; (D) New Zealand; (E) Philippines.

### 3.4 Insights from the PDF

#### 3.4.1 PDF for displacement distribution

We can utilize the PDF, Equation (3), to gain added insights into earthquake events for a given seismic event interval, for example at $\tau = 100$. With the displacement, $\Delta x = (x_1 - x_0)$, where $x$ can be the earthquake epicenter distance, depth, or magnitude, we generate the PDF in Figure 10. For the case of earthquake epicenter distances, the peak of the PDF decreases while the range of displacement values slightly spreads as lower magnitude events are filtered out as seen in Figure 10A. This shows that sequences of higher magnitude earthquake events are less likely and are spatially distant from each other. Similarly for earthquake depths, Figure 10B illustrates that the PDF of consecutive seismic events increases and exhibits less spreading in the displacement values when cutoff magnitude is decreased. This indicates that consecutive high magnitude earthquakes are less likely to cluster in depth compared to when lower magnitude seismic events are taken into account. Consequently, removal of lower magnitude earthquakes narrows the range of the possible displacement values while increasing the peak PDF of the magnitude dataset as shown in Figure 10C. A similar behavior is observed when other event intervals and regions are used.

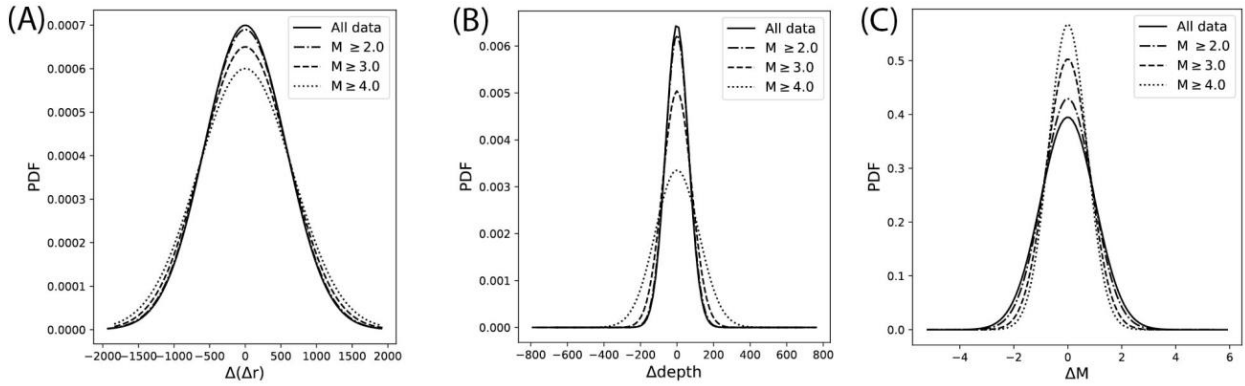

**Figure 10. PDF of earthquake (A) interevent distances, (B) depths, and (C) magnitudes from Equation (3) and using the datasets in Figure 4 for $\tau$=100.**

### 3.4.2 Earthquake recurrence

The temporal evolution of the PDF for the earthquake epicenter distances, depths, and magnitudes can also be generated using Equation (3) as a function of the occurrence interval $\tau$ for different magnitude cutoffs. Using the obtained $\beta$ and $\mu$ values ( Supplemental Material, Note S1), the plot of PDF versus occurrence interval $\tau$ is shown in Figure 11. For all the three datasets extracted from the catalog, the PDF of seismic event recurrence decreases with increasing magnitude cutoff and as time progresses. The decrease is seen to be more pronounced at shorter intervals. Figure 11 indicates that the probability of earthquake recurrence, that is an earthquake occurring at the same epicenter (Figure 11A), depth (Figure 11B), or with the same magnitude (Figure 11C), is higher at smaller $\tau$. Looking at Figure 11A and 11B, when lower magnitude events are disregarded, the probability of earthquake recurrence decreases. Results confirm that recurrence of higher magnitude events at the same epicenter (Figure 11A) or depth (Figure 11B) are less likely than when lower magnitude earthquakes are taken into consideration. Figure 11C illustrates that recurrence of earthquakes with the same magnitude increases as the more frequent (lower magnitude) seismic events are not taken into account.

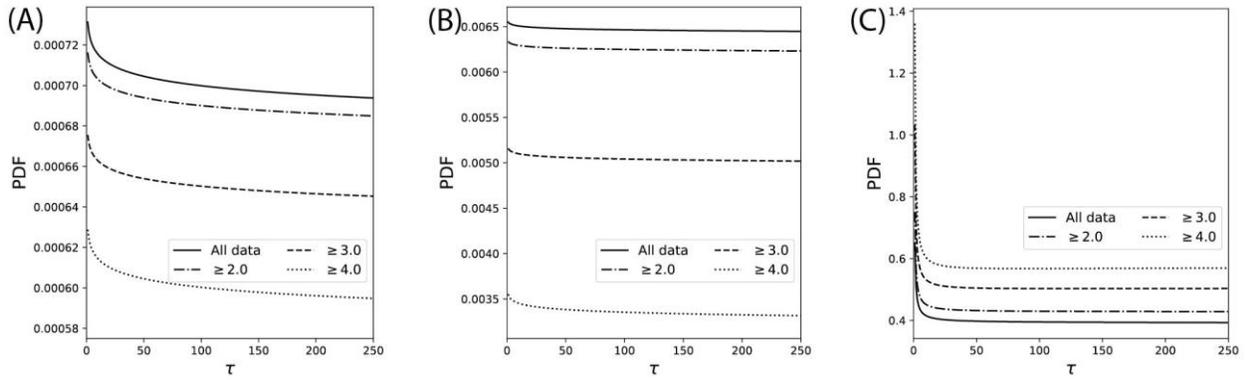

Figure 11. Recurrence PDF of earthquake (A) interevent distances, (B) depths, and (C) magnitudes from Equation (3) and using the datasets in Figure 5.

### 3.5 Insights from parameters $\mu$ and $\beta$

Crucial to matching the theoretical and empirical MSD and PDF plots are the values of the parameters $\mu$ and $\beta$ in the analytical model given by Equations (2) to (4). These parameters provide added insights to the memory behavior and dynamics of the system involved. Highly noticeable is the relatively constant value of the memory parameter $\mu$ regardless of whether convergent, divergent, or transform plate boundaries are involved in different regions of the Pacific Ring of Fire (Figure 1). For example, for All Data, the various values of $\mu$ (see Table 2) ranged only between 1.019 and 0.984 for all cases of epicenter distances, depth, and magnitude. The almost constant value of $\mu$ strongly shows a collective underlying universal memory behavior for earthquake epicenter distances, depths, and magnitudes notwithstanding the different locations. This is very unlike another system, such as the gelling of fibrin (Aure et al. 2019), which also used Equations (2) to (4). In fibrin, the parameter $\mu$ decreased dramatically as fibrin ages. We note however that for earthquakes, the relatively constant value of $\mu$ could arise from the fact that the earthquake datasets were taken between 1994 and 2021 which is an extremely short time interval – a snapshot of earth's history – compared to the age of earth. It is possible that datasets very far removed from these dates may exhibit a different value of $\mu$, but still in a collective manner, i.e, same value $\mu$ across different regions in the Pacific Ring of Fire implying the same memory behavior for epicenter distances, depths, and magnitudes for a given interval of time.

The decay parameter $\beta$, on the other hand, is in stark contrast to the generally constant value of $\mu$. The $\beta$ exhibits location-dependent changing values for the earthquake interevent distances, depths, and magnitudes. Several observations, however, can be pointed out. For earthquake epicenter distances, Japan and the Philippines generate relatively close values for the decay parameter $\beta$ (see, e.g., Figure 6A and the Supplemental Material, Note S1). These close values in $\beta$ could possibly arise from the fact that the motion and subduction of the Philippine plate beneath the Eurasian plate is directed both to the Philippines and Japan (Figure 1). Another noticeable similarity is that running across both Philippines and Japan are very long strike slip fault systems – the 1,200 km long Philippine fault (Aurelio and Peña, 2010) along its north-south axis and Japan's longest fault system, the Median Tectonic Line (Tsutsumi and Okada, 1996).

This stochastic study also reveals that earthquakes in Mexico are markedly different from others. For epicenter distances, depths, and magnitudes, the $\beta$ values of Mexico by and large deviate from that of other countries. For earthquake magnitudes, Mexico's values of the decay parameter $\beta$ are way above other countries (Table 2), while its MSD is very low in comparison. A possible explanation could come from the fact that, in Mexico, all three types of plate boundaries interact: convergent (subduction), divergent (spreading center), and transform plate boundaries (Figure 1). Mexico is also in the intersection of four plates: the Cocos Plate, Pacific Plate, Caribbean Plate, and the North American Plate.

For Southern California, its decay parameter $\beta$ for All Data is also characteristically different from other countries in epicenter distances, depth, and magnitude. This different values of $\beta$ for Southern California may be due to the fact that it lies uniquely in a transform plate boundary between the Pacific plate and the North American plate (Figure 1). This is unlike other countries where a convergent plate boundary is also involved. Transform plate boundaries are normally associated with shallow depth earthquakes of large magnitude.

The countries Chile and New Zealand, though located far from each other, exhibit close values for the decay parameter $\beta$ as far as earthquake magnitudes for the entire data set are concerned (see, e.g., Figure 6C). This could possibly be due to both countries having convergent and transform plate boundaries (Figure 1). A closer look, however, shows that their $\beta$ values diverge for different cut-off magnitudes, i.e., $\geq 2.0, \geq 3.0, \geq 4.0$ (see, e.g. Supplemental Material, Note S1). It has been noted that, the very long north-south stretch on the western coast of Chile experiences one of the fastest rates of convergence at 66mm per year between the Nazca plate and South American plate (Maksymowicz, 2015). These two plates are joined only at the southernmost tip of Chile− the Scotia plate and Antarctic plate.

## 4      Conclusion

Earthquake occurrences have been observed to possess memory. In this paper, using white noise analysis, we modeled decades-long of earthquake dataset for regions in the tectonically active Pacific Ring of Fire. We analyzed the dataset for distances between consecutive earthquake epicenters and depths and showed that these exhibit a stochastic process with memory described by Equations (2)-(4). Successive earthquake epicenters and hypocenter depths are re-imagined as trajectories of a single diffusive point where one can extract an MSD for the fluctuating positions. The MSD indicates that the 'motion' of earthquakes is generally highly sub-diffusive. We also model the dataset of earthquake magnitudes using the same set of functions. There is an apparent clustering of seismic events depending on the various cutoff magnitudes. Earthquake epicenter distances, depths, and magnitudes for the regions in the Pacific Ring of Fire are collectively described by the PDF given by Equation 3. From the PDF, we confirm that seismic events (recurrence or not) of higher magnitudes are less likely, which is consistent with the Gutenberg-Richter law. The derived PDF and MSD functions are also found to be insensitive to earthquake catalog completeness which provides a more versatile approach in analyzing dataset series obtained from earthquake catalogs.

**Conflict of interest**

The authors declare that the research was conducted in the absence of any commercial or financial relationships that could be construed as a potential conflict of interest.

**Author Contributions**

PJR, RRV, and CCB conceptualized the paper. PJR curated the data and generated all the plots. PJR and CCB wrote the paper. CCB and RRV supervised technical input in generating the graphs and analysis. JS provided data visualization, graphics, and technical input of geological explanations. All the authors contributed to the editing, presentation, and reviewing of the manuscript.

**Acknowledgments**

The financial support given by the Department of Science and Technology – Science Education Institute (DOST-SEI) Accelerated Science and Technology Human Resource Development Program (ASTHRDP) is acknowledged. The seismological data were obtained by the institutions listed in Table 1. The authors thank all the staff for the maintenance of the stations, the acquisition and distribution of the data.

Zhang, Y., Ashkenazy, Y., Havlin, S. (2021b). Asymmetry in Earthquake Interevent Time Intervals. Journal of Geophysical Research: Solid Earth, 126(9). https://doi.org/10.1029/2021JB022454

**Data availability statement**
Publicly available datasets were analyzed in this study. The sources of these datasets can be found in the article.

# *Supplementary Material*

## Earthquake Occurrences in the Pacific Ring of Fire Exhibit a Collective Stochastic Memory for Magnitudes, Depths, and Relative Distances of Events

**Pamela Jessica C. Roque*, Renante R. Violanda, Christopher C. Bernido and Janneli Lea A. Soria**

### S1  Parameter values

The complete list of β and μ values that provide the best fit for Equation (4) and the empirical data for earthquake epicenter distances, depths, and magnitudes for the different regions in the Pacific Ring of Fire is provided below. The parameter values for various magnitude cutoffs for each region are listed as well.

| All Data | β | | | μ | | |
|---|---|---|---|---|---|---|
| | Epicenter | Depth | Magnitude | Epicenter | Depth | Magnitude |
| Chile | $(2.250 \pm 0.05) \times 10^{-6}$ | $(1.916 \pm 0.003) \times 10^{-4}$ | $1.195 \pm 0.003$ | $1.0149 \pm 0.0001$ | $1.0154 \pm 0.0001$ | $1.0004 \pm 4.52 \times 10^{-4}$ |
| Japan | $(3.839 \pm 0.007) \times 10^{-6}$ | $(3.9728 \pm 0.0009) \times 10^{-4}$ | $0.825 \pm 0.002$ | $1.00676 \pm 0.00009$ | $1.00193 \pm 0.00002$ | $0.9998 \pm 0.0003$ |
| Mexico | $(1.679 \pm 0.005) \times 10^{-6}$ | $(4.014 \pm 0.003) \times 10^{-4}$ | $3.74 \pm 0.02$ | $1.0094 \pm 0.0002$ | $1.00444 \pm 0.00005$ | $0.9835 \pm 0.0009$ |
| New Zealand | $(6.69 \pm 0.02) \times 10^{-6}$ | $(1.947 \pm 0.001) \times 10^{-4}$ | $1.187 \pm 0.003$ | $1.0129 \pm 0.0001$ | $1.00516 \pm 0.00004$ | $1.0129 \pm 0.0005$ |
| Philippines | $(4.21 \pm 0.01) \times 10^{-6}$ | $(2.813 \pm 0.004) \times 10^{-4}$ | $0.990 \pm 0.002$ | $1.0191 \pm 0.0001$ | $1.00563 \pm 0.00009$ | $1.0191 \pm 0.0004$ |
| Southern California | $(4.482 \pm 0.008) \times 10^{-5}$ | $(2.822 \pm 0.001) \times 10^{-2}$ | $1.529 \pm 0.005$ | $1.0157 \pm 0.0001$ | $1.01202 \pm 0.00004$ | $1.0157 \pm 0.0005$ |

| Magnitudes $\geq 2.0$ | β | | | μ | | |
|---|---|---|---|---|---|---|
| | Epicenter | Depth | Magnitude | Epicenter | Depth | Magnitude |
| Chile | $(2.256 \pm 0.005) \times 10^{-6}$ | $(1.931 \pm 0.004) \times 10^{-4}$ | $1.240 \pm 0.003$ | $1.0151 \pm 0.0001$ | $1.0161 \pm 0.0001$ | $0.9993 \pm 0.0005$ |
| Japan | $(6.85 \pm 0.02) \times 10^{-6}$ | $(1.2906 \pm 0.0009) \times 10^{-4}$ | $1.383 \pm 0.004$ | $1.0250 \pm 0.0001$ | $1.00590 \pm 0.00005$ | $0.9953 \pm 0.0005$ |
| Mexico | $(1.678 \pm 0.005) \times 10^{-6}$ | $(4.010 \pm 0.003) \times 10^{-4}$ | $3.90 \pm 0.02$ | $1.0094 \pm 0.0002$ | $1.00443 \pm 0.00005$ | $0.9825 \pm 0.0009$ |
| New Zealand | $(5.89 \pm 0.01) \times 10^{-6}$ | $(1.609 \pm 0.001) \times 10^{-4}$ | $1.667 \pm 0.005$ | $1.0150 \pm 0.0001$ | $1.00699 \pm 0.00004$ | $1.0043 \pm 0.0006$ |
| Philippines | $(3.91 \pm 0.01) \times 10^{-6}$ | $(2.645 \pm 0.004) \times 10^{-4}$ | $1.138 \pm 0.003$ | $1.0162 \pm 0.0002$ | $1.0060 \pm 0.0001$ | $0.9979 \pm 0.0004$ |
| Southern California | $(4.20 \pm 0.01) \times 10^{-5}$ | $(3.182 \pm 0.003) \times 10^{-2}$ | $2.391 \pm 0.009$ | $1.0399 \pm 0.0002$ | $1.0170 \pm 0.0001$ | $0.9964 \pm 0.0007$ |

| Magnitudes ≥ 3.0 | β | | | μ | | |
|---|---|---|---|---|---|---|
| | Epicenter | Depth | Magnitude | Epicenter | Depth | Magnitude |
| Chile | $(2.515 \pm 0.009) \times 10^{-6}$ | $(1.800 \pm 0.004) \times 10^{-4}$ | $1.638 \pm 0.005$ | $1.0206 \pm 0.0002$ | $1.0218 \pm 0.0002$ | $0.9940 \pm 0.0006$ |
| Japan | $(1.025 \pm 0.003) \times 10^{-5}$ | $(8.69 \pm 0.01) \times 10^{-5}$ | $1.792 \pm 0.006$ | $1.0412 \pm 0.0002$ | $1.00667 \pm 0.00009$ | $0.9925 \pm 0.0006$ |
| Mexico | $(1.776 \pm 0.006) \times 10^{-6}$ | $(3.952 \pm 0.003) \times 10^{-4}$ | $4.40 \pm 0.02$ | $1.0095 \pm 0.0002$ | $1.00435 \pm 0.00006$ | $0.9755 \pm 0.0009$ |
| New Zealand | $(5.76 \pm 0.02) \times 10^{-6}$ | $(1.236 \pm 0.003) \times 10^{-4}$ | $2.308 \pm 0.008$ | $1.0331 \pm 0.0002$ | $1.0276 \pm 0.0002$ | $0.9943 \pm 0.0006$ |
| Philippines | $(3.50 \pm 0.01) \times 10^{-6}$ | $(1.808 \pm 0.004) \times 10^{-4}$ | $1.504 \pm 0.004$ | $1.0166 \pm 0.0002$ | $1.0099 \pm 0.0001$ | $0.9888 \pm 0.0005$ |
| Southern California | $(9.3 \pm 0.1) \times 10^{-5}$ | $(3.68 \pm 0.02) \times 10^{-2}$ | $2.55 \pm 0.01$ | $1.118 \pm 0.001$ | $1.0425 \pm 0.0005$ | $0.9842 \pm 0.0007$ |

| Magnitudes ≥ 4.0 | β | | | μ | | |
|---|---|---|---|---|---|---|
| | Epicenter | Depth | Magnitude | Epicenter | Depth | Magnitude |
| Chile | $(2.67 \pm 0.01) \times 10^{-6}$ | $(1.863 \pm 0.009) \times 10^{-4}$ | $2.341 \pm 0.009$ | $1.0323 \pm 0.0003$ | $1.0482 \pm 0.0003$ | $0.9942 \pm 0.0007$ |
| Japan | $(8.21 \pm 0.99) \times 10^{-6}$ | $(1.452 \pm 0.006) \times 10^{-4}$ | $2.184 \pm 0.008$ | $1.0289 \pm 0.0007$ | $1.0074 \pm 0.0003$ | $1.0020 \pm 0.0007$ |
| Mexico | $(1.98 \pm 0.01) \times 10^{-6}$ | $(2.938 \pm 0.006) \times 10^{-4}$ | $5.02 \pm 0.02$ | $1.0189 \pm 0.0003$ | $1.0090 \pm 0.0001$ | $0.964 \pm 0.001$ |
| New Zealand | $(4.47 \pm 0.03) \times 10^{-6}$ | $(9.60 \pm 0.06) \times 10^{-5}$ | $3.01 \pm 0.01$ | $1.0341 \pm 0.0004$ | $1.0337 \pm 0.0004$ | $0.9740 \pm 0.0008$ |
| Philippines | $(3.17 \pm 0.02) \times 10^{-6}$ | $(9.85 \pm 0.06) \times 10^{-5}$ | $1.858 \pm 0.007$ | $1.0201 \pm 0.0003$ | $1.0250 \pm 0.0004$ | $0.9802 \pm 0.0007$ |
| Southern California | – | – | – | – | – | – |

## S2 Probability density function with memory

The fluctuating earthquake variable $x$ starts at $x_0$ as parametrized by Equation (2) can be anywhere at some later time $t$. We can pin down the final point such that at final time $t$ we have, $x(t) = x_1$, using the Donsker delta function (Hida et al. 1993), i.e.,

$$\delta(x(t) - x_1) = \delta\left(x_0 + \int_0^t (t-\tau)^{(\mu-1)/2} \frac{exp(-\beta/2\tau)}{\tau^{(\mu+1)/2}} dB(\tau) - x_1\right), \quad (S1)$$

In Equation (S1), we we have written $x(t)$ explicitly using Equation (2). For all fluctuating paths that do not end at $x_1$, the delta function Equation (S1) vanishes. Note that Equation (S1) can be written as a functional of the white noise random variable $\omega(\tau)$ (Hida et al 1993) where, $\omega(\tau) = dB(\tau)/d\tau$. One can then obtain the probability density function, $P(x_1, t; x_0, 0)$, that a path ends at $x_1$ if it started at $x_0$, by integrating over all possible paths that satisfy Equation (S1), i.e.,

$$P(x_1, t; x_0, 0) = \int \delta(x(t) - x_1) \, d\mu(\omega), \quad (S2)$$

where $d\mu(\omega)$ is the Gaussian white noise measure (Hida et al 1993). Expressing the delta function in terms of its Fourier representation we have,

$$P(x_1, t; x_0, 0) = \frac{1}{\sqrt{2\pi}} \iint_{-\infty}^{+\infty} e^{ik(x(t)-x_1)} dk \, d\mu(\omega)$$

$$= \frac{1}{\sqrt{2\pi}} \int_{-\infty}^{+\infty} dk \, e^{ik(x_0-x_1)} \int e^{ik\left[\int_0^t (t-\tau)^{(\mu-1)/2} \frac{exp(-\beta/2\tau)}{\tau^{(\mu+1)/2}} \omega(\tau) d\tau\right]} d\mu(\omega)$$

$$= \frac{1}{\sqrt{2\pi}} \int_{-\infty}^{+\infty} dk \, e^{ik(x_0-x_T)} \int e^{i\int_0^t \omega(\tau)\xi(\tau)d\tau} d\mu(\omega) , \quad \text{(S3)}$$

where we let, $\xi(\tau) = k(t-\tau)^{(\mu-1)/2} exp(-\beta/2\tau)/\tau^{(\mu+1)/2}$. Integration over $d\mu(\omega)$ is done using the characteristic functional (Hida et al. 1993),

$$\int e^{i\int_0^t \omega(\tau)\xi(\tau)d\tau} d\mu(\omega) = e^{-\frac{1}{2}\int_0^t \xi(\tau)^2 d\tau} , \quad \text{(S4)}$$

to get,

$$P(x_1, t; x_0, 0) = \frac{1}{\sqrt{2\pi}} \int_{-\infty}^{+\infty} e^{ik(x_0-x_1) - \frac{k^2}{2}\left[\int_0^t t-\tau^{\mu-1} exp(-\beta/\tau)/\tau^{\mu+1} d\tau\right]} dk , \quad \text{(S5)}$$

The integral over $dk$ is a Gaussian integral which yields $P(x_1, t; x_0, 0)$ given by Equation (3).

### References

Hida, T., Kuo, H.H., Potthoff, J., and Streit, L. (2013). White noise: an infinite dimensional calculus (Vol. 253). Springer Science & Business Media.

Bernido, C.C. and Carpio-Bernido, M.V. (2015). *Methods and Applications of White Noise Analysis in Interdisciplinary Sciences*. Singapore: World Scientific.

Bernido, C.C. and Carpio-Bernido, M.V. (2012). White noise analysis: some applications in complex systems, biophysics and quantum mechanics. *Int. J. Mod. Phys. B, 26*(12300014).